\documentclass[prl,aps,floatfix,showpacs,twocolumn]{revtex4}
\usepackage{dcolumn}
\usepackage{bm}
\usepackage{graphicx}
\begin{document}

\title{Enhanced effect of quark mass variation in $^{229}$Th
  and limits from Oklo data}
\author{V. V. Flambaum$^{1,3,4}$ and R. B. Wiringa$^2$}
\affiliation{
$^1$\mbox{Argonne Fellow, Physics Division, Argonne National Laboratory,
           Argonne, Illinois 60439} \\
$^2$\mbox{Physics Division, Argonne National Laboratory, 
           Argonne, Illinois 60439}\\
$^3$\mbox{School of Physics, University of New South Wales, Sydney 2052,
           Australia} \\
$^4$\mbox{Perimeter Institute for Theoretical Physics, 
           Waterloo, Ontario N2L 2Y5, Canada}
}

\date{\today}

\begin{abstract}
The effects of the variation of the dimensionless strong interaction parameter
$X_q=m_q/\Lambda_{QCD}$ ($m_q$ is the quark mass, $\Lambda_{QCD}$ is the QCD
scale) are enhanced about $1.5 \cdot 10^5$ times in the 7.6 eV 
``nuclear clock'' transition between the ground and first excited states 
in the $^{229}$Th nucleus and about $1 \cdot 10^8$ times in the relative 
shift of the 0.1 eV compound resonance in $^{150}$Sm.
The best terrestrial limit on the temporal variation of the fundamental 
constants,
 $|\delta X_q/X_q| < 4 \cdot 10^{-9} $ at 1.8 billion years ago
($|{\dot X}_q/X_q|<2.2 \cdot 10^{-18}$ y$^{-1}$) , 
is obtained from the shift of this Sm resonance derived from the Oklo 
natural nuclear reactor data.
The results for  $^{229}$Th and $^{150}$Sm are obtained by extrapolation
from light nuclei where the many-body calculations can be performed more 
accurately.
The errors produced by such extrapolation may be
smaller than the errors of direct calculations in heavy nuclei.
The extrapolation results are compared with the ``direct'' estimates obtained
using the Walecka model.
A number of numerical relations needed for the calculations
of the variation effects in nuclear physics and atomic spectroscopy
have been obtained: for the nuclear binding energy
$\delta E/E \approx -1.45~\delta m_q/m_q$, for the spin-orbit intervals
$\delta E_{so}/E_{so} \approx -0.22~\delta m_q/m_q$,
for the nuclear radius $\delta r/r \approx 0.3~\delta m_q/m_q$
 (in units of $\Lambda_{QCD}$); 
 for the shifts of nuclear
 resonances and weakly bound energy levels
$\delta E_r \approx 10~\delta X_q/X_q$  MeV.  
\end{abstract}
\pacs{PACS: 06.20.Jr, 42.62.Fi, 23.20.-g}
\maketitle

\section{Introduction}

Unification theories applied to cosmology suggest the possibility
of variation of the fundamental constants in
the expanding Universe (see, e.g., the review~\cite{Uzan}). 
A review of recent results can be found, e.g., in Ref.~\cite{F2007}.
In  Ref.~\cite{th1} it was suggested that there may
be a five orders of magnitude enhancement of the variation effects
in the low-energy transition between the
ground and the first excited states in the $^{229}$Th nucleus.
This transition was suggested as a possible nuclear clock in Ref.~\cite{th4}.
Indeed, the transition is very narrow.  
The width of the excited state is estimated to be about 
$10^{-4}$ Hz~\cite{th2}. 
The latest measurement of the transition energy~\cite{Beck} gives
$7.6 \pm 0.5$ eV, compared to earlier values of $5.5\pm 1$ eV~\cite{th6} 
and $3.5\pm 1$ eV~\cite{th}.
Therefore, this transition may be investigated using laser spectroscopy 
where the relative accuracy has already reached $10^{-16}$. 
Several experimental groups have already started working on this 
possibility~\cite{exp}. 
However, a recent paper~\cite{Hayes} claims that there is not any
enhancement of the effects of the variation of the fundamental constants 
in this transition. 
The main aim of the present note is to demonstrate that the enhancement exists. 
We also estimate the relative shift of the 0.1 eV compound resonance 
in $^{150}$Sm to obtain new limits on the variation of the fundamental 
constants from the Oklo natural nuclear reactor data \cite{Gould,Petrov,Fujii}.

We can measure only the variation of dimensionless parameters which do not
depend on which units we use. 
In the Standard Model, the two most important dimensionless parameters are
the fine structure constant $\alpha=e^2/\hbar c$ and the ratio
of the electroweak unification scale determined by the Higgs
vacuum expectation value (VEV) to the quantum chromodynamics
(QCD) scale $\Lambda_{QCD}$ (defined as the position of the Landau pole in
the logarithm for the running strong coupling constant,
$\alpha_s(r) \sim$ constant$/\ln{(\Lambda_{QCD} r/\hbar c)}$). 
The variation of the Higgs VEV leads to the variation of the fundamental 
masses which are proportional to the Higgs VEV.
The present work considers mainly effects produced by the variation of
$X_q=m_q/\Lambda_{QCD}$ where $m_q=(m_u+m_d)/2$ is the average light quark
mass. 
Within Grand Unification Theories the relative variation of $X_q$ may 
be 1--2 orders of magnitude larger than the variation of 
$\alpha$~\cite{Marciano}. 
Note that in the present work we do not consider effects of variation of 
the strange quark mass since they have larger uncertainty and should be 
treated separately.
These effects were estimated in Refs.~\cite{Shuryak2003,th1}.

The results depend on the dimensionless parameter  $X_q=m_q/\Lambda_{QCD}$.
In all calculations
it is convenient to assume that $\Lambda_{QCD}$ is constant and calculate
the dependence on the small parameter $m_q$. In other words, we measure
all masses and energies in units of $\Lambda_{QCD}$ and will simply restore 
$\Lambda_{QCD}$ in the final results. Note that when a relative effect
of the variation is enhanced it does not matter what units we use.
The variation of the ratio of different units may be neglected anyway.
 
\section{Thorium}

To explain the origin of the enhancement we should  present
the small 7.6 eV interval between the ground and excited states
in the $^{229}$Th nucleus as a sum of a few components which nearly cancel
each other and have different dependence on the fundamental constants.
If one performs the calculations exactly, it does not matter
how we select these components. However, in practice the calculations
are always approximate, therefore, a reasonable selection of the
components will determine our final accuracy.
For example, to study dependence on $\alpha$ we should separate
the Coulomb energy from the remaining contributions to the energy.
To study dependence on $X_q=m_q/\Lambda_{QCD}$ it is convenient 
to separate out the spin-orbit interaction energy:
\begin{equation}\label{omega}
  \omega=E_b + E_{so}=7.6~{\rm eV} \ .
\end{equation}
Here $E_b$ is the difference in bulk binding energies of the excited and 
ground states (including kinetic and potential energy but excluding the 
spin-orbit interaction) and $E_{so}$ is the difference in the spin-orbit
interaction energies $V_{ls} \langle{\bf l \cdot s}\rangle$ in the excited 
and ground states.
We make this separation because we expect $E_b$ and $E_{so}$ to have
a very different dependence on $X_q=m_q/\Lambda_{QCD}$, as discussed below.
In $^{229}$Th the strength of the spin-orbit interaction is estimated to be
$V_{ls}= -0.85$ MeV from Table 5-1 of Ref.~\cite{BM}.
The difference of $\langle{\bf l \cdot s}\rangle$ between the excited and
ground states can be easily calculated using the expansion of the wave
functions over Nilsson orbitals presented in Table 4 of Ref.~\cite{Gulda}:
$E_{so}\approx 1.22 V_{ls}\approx -1.04$ MeV.
(Note that without configuration mixing, for the ``pure'' Nilsson
excited state $[631]3/2^+$ and ground state $[633]5/2^+$, $E_{so}=2 V_{ls}$.)
Then Eq.~(\ref{omega}) gives us $E_b\approx -E_{so}\approx 1$ MeV and
\begin{equation}\label{delta1}
  \frac{ \delta \omega}{\omega}\approx\frac{E_{so}}{\omega}
(\frac{ \delta E_{so}} {E_{so}}-\frac{\delta E_b}{E_b})=1.3 \cdot 
10^5~(\frac{\delta E_{so}}{E_{so}}-\frac{\delta E_b}{E_b}) \ .
\end{equation}

Qualitatively, we expect $E_b$ and $E_{so}$ to have a rather different
dependence on $X_q$.
In the Walecka model (which was used in Ref.~\cite{th1} to 
estimate the enhancement factor) there is a significant cancellation 
between the $\sigma$ and $\omega$ meson contributions to the mean-field 
potential and the total binding energy $E$, while the $\sigma$ and $\omega$
mesons contribute with equal sign to the spin-orbit interaction constant 
$V_{ls}$~\cite{C}.
A similar argument may be made from the variational Monte Carlo (VMC)
calculations with realistic interactions used in Ref.~\cite{FW07}
to evaluate binding energy dependence on $X_q$.
These calculations use nucleon-nucleon potentials that fit $N\!N$ scattering 
data together with three-nucleon potentials that reproduce the binding 
energies of light nuclei, 
The binding energy is the result of a significant cancellation between 
intermediate-range attraction due to two-pion exchange and short-range 
repulsion arising from heavy vector-meson exchange.  
However, spin-orbit splitting between nuclear levels has been found to 
be a coherent addition of short-range two-nucleon ${\bf l \cdot s}$
interaction and multiple-pion exchange between three or more 
nucleons \cite{PP93}.  
Thus if meson masses move in the same direction due to an underlying 
quark mass shift, contributions from pion exchange and heavy vector-meson 
exchange will cancel against each other in the binding energy, but reinforce
each other in spin-orbit splittings.

\subsection{Binding energies}

The binding energy per nucleon and the spin-orbit interaction constant
have a slow dependence on the nucleon number $A$. 
The total binding is dominated by the bulk terms, so we make the 
reasonable assumption that the variation of the bulk energy with $X_q$ 
is the same for the two levels in $^{229}$Th and thus the variation of the
difference $\delta E_b/E_b \approx \delta E/E$.
Moreover, the common factors (like $A^{-1/3}$ in the spin-orbit constant 
$V_{ls}$ \cite{BM}) cancel out in the relative variations
$\delta E_{so}/ E_{so}$ and $\delta E_b/ E_b$. 
Therefore, it may be plausible to extract these relative variations from 
the type of calculations in light nuclei performed in Ref.~\cite{FW07}. 
The advantage of the light nuclei is that the calculations can be performed 
quite accurately, including different many-body effects. 
Their accuracy has been tested by comparison with the experimental data for
the binding energies and by comparison of the results obtained using several
sophisticated interactions (AV14, AV28, AV18+UIX -- see~\cite{FW07}). 
As the first step, the variations of the nuclear binding energies have been 
expressed in terms of the variations of nucleon, $\Delta$, pion and 
vector-meson masses.
The dependence of these masses on quark masses have been taken from
Refs.~\cite{FHJRW06,HMRW}. 
The results for the relative variations of the total binding energies
are presented in Table~\ref{tab:mq} (in the present work we add 
$^6$He, $^7$He, and $^9$Be to this table).
We see that all the results are close to the average value
$\delta E/E \approx -1.45~\delta X_q/X_q$. 
The maximal deviations are for $^4$He, which is especially 
tightly bound, and for $^7$He, which is a resonant state.

\begin{table*}[ht!]
\caption{Dimensionless derivatives
  $K = \frac{\delta E/E}{\delta X_q/X_q}$ of the binding energy over
  $X_q=m_q/\Lambda_{QCD}$ . }
\begin{ruledtabular}
\begin{tabular}{rrrrrrrrrrrr}
\multicolumn{1}{c}{$^2$H}&\multicolumn{1}{c}{$^3$H}
&\multicolumn{1}{c}{$^3$He}&\multicolumn{1}{c}{$^4$He}
&\multicolumn{1}{c}{$^5$He}&\multicolumn{1}{c}{$^6$He}
&\multicolumn{1}{c}{$^6$Li}&\multicolumn{1}{c}{$^7$He}
&\multicolumn{1}{c}{$^7$Li}&\multicolumn{1}{c}{$^7$Be}
&\multicolumn{1}{c}{$^8$Be}&\multicolumn{1}{c}{$^9$Be}\\
\colrule
$-1.39$ & $-1.44$ & $-1.55$ & $-1.08$ & $-1.24$ & $-1.50$ &
$-1.36$ & $-1.93$ & $-1.50$ & $-1.57$ & $-1.35$ & $-1.59$ \\
\end{tabular}
\end{ruledtabular}
\label{tab:mq}
\end{table*}

\subsection{Spin-orbit intervals}

To find the dependence of the spin-orbit constant $V_{ls}$ on
$m_q/\Lambda_{QCD}$ we calculate the spin-orbit splitting between the 
$p_{1/2}$ and $p_{3/2}$ levels in $^5$He, $^7$He, $^7$Li, and $^9$Be in the 
present work.
We use the Argonne v$_{18}$ two-nucleon and Urbana IX three-nucleon
(AV18+UIX) interaction which provides our best results for small nuclei
(see Ref.~\cite{FW07} for details and references).
In all calculations it is convenient to keep $\Lambda_{QCD}=$ constant, 
i.e., measure the quark mass $m_q$ in units of $\Lambda_{QCD}$.
We restore $\Lambda_{QCD}$ in the final answers.
As the first step we calculate the binding energies of the ground and
excited states shown in Table~\ref{tab:energy} and their dependence 
on the nucleon, $\Delta$, pion, and vector-meson masses,
$\Delta{\mathcal E}(m_H) = \frac{\delta E/E}{\delta m_H/m_H}$,
shown in Table~\ref{tab:av18uix}.
To find the dependence of these energies on the quark mass, we utilize the
results of a Dyson-Schwinger equation (DSE) study of sigma terms in
light-quark hadrons~\cite{FHJRW06}.  
Equations (85-86) of that work give the rate of hadron mass variation as 
a function of the average light current-quark mass 
$m_q = (m_u+m_d)/2$ as:
\begin{equation}
  \frac{\delta m_H}{m_H}
   = \frac{\sigma_H}{m_H} \frac{\delta m_q}{m_q} \ ,
\label{eq:dse}
\end{equation}
with $\sigma_H/m_H$ values of 0.498 for the pion, 0.030 for the $\rho$-meson,
0.043 for the $\omega$-meson, 0.064 for the nucleon, and 0.041 for the 
$\Delta$.
The values for the $\rho$ and $\omega$-mesons were reduced to 0.021 and 
0.034 in subsequent work~\cite{HMRW}.  
We use an average of the $\rho$ and $\omega$ terms of 0.030 for our 
short-range mass parameter $m_V$.

\begin{table*}[ht!]
\caption{Experimental and calculated energies for the ground
($p_{3/2}$) and first excited ($p_{1/2}$) states of $A$=5,7,9 nuclei in MeV.}
\begin{ruledtabular}
\begin{tabular}{lrrrrrrrr}
& \multicolumn{1} {c}{$^5$He($\frac{3}{2}^-$)} 
& \multicolumn{1} {c}{$^5$He*($\frac{1}{2}^-$)}
& \multicolumn{1} {c}{$^7$He($\frac{3}{2}^-$)} 
& \multicolumn{1} {c}{$^7$He*($\frac{1}{2}^-$)}
& \multicolumn{1} {c}{$^7$Li($\frac{3}{2}^-$)} 
& \multicolumn{1} {c}{$^7$Li*($\frac{1}{2}^-$)}
& \multicolumn{1} {c}{$^9$Be($\frac{3}{2}^-$)} 
& \multicolumn{1} {c}{$^9$Be*($\frac{1}{2}^-$)}\\
\colrule
 AV18+UIX  & $-25.26$ & $-24.02$ & $-21.77$ & $-19.56$
           & $-33.33$ & $-33.02$ & $-45.39$ & $-42.01$ \\
 Expt.     & $-27.41$ & $-26.23$ & $-28.83$ & $-26.23$
           & $-39.24$ & $-38.77$ & $-58.16$ & $-55.38$ \\
\end{tabular}
\end{ruledtabular}
\label{tab:energy}
\end{table*}

\begin{table*}[ht!]
\caption{Dimensionless derivatives
$\Delta{\mathcal E}(m_H) = \frac{\delta E/E}{\delta m_H/m_H}$ of the
binding energy to the different hadron masses and the 
sensitivity $K$ after folding in the DSE values of $\delta m_H/m_H$.}
\begin{ruledtabular}
\begin{tabular}{lrrrrrrrr}
& \multicolumn{1} {c}{$^5$He($\frac{3}{2}^-$)} 
& \multicolumn{1} {c}{$^5$He*($\frac{1}{2}^-$)}
& \multicolumn{1} {c}{$^7$He($\frac{3}{2}^-$)} 
& \multicolumn{1} {c}{$^7$He*($\frac{1}{2}^-$)}
& \multicolumn{1} {c}{$^7$Li($\frac{3}{2}^-$)} 
& \multicolumn{1} {c}{$^7$Li*($\frac{1}{2}^-$)}
& \multicolumn{1} {c}{$^9$Be($\frac{3}{2}^-$)} 
& \multicolumn{1} {c}{$^9$Be*($\frac{1}{2}^-$)}\\
\colrule
$m_N+\delta_N$  &  13.31  &  13.83  &  19.34  &  21.34
                &  15.53  &  15.48  &  16.09  &  17.12 \\
$\delta_\Delta$ &$-10.24$ &$-10.72$ &$-14.92$ &$-16.63$
                &$-11.96$ &$-11.88$ &$-12.39$ &$-13.27$\\
$m_\pi$ (+TNI)  & $-5.82$ & $-6.07$ & $-8.78$ & $-9.73$
                & $-6.91$ & $-6.88$ & $-7.27$ & $-7.76$\\
$m_V$           &  40.87  &  42.84  &  60.46  &  67.54
                &  48.11  &  47.81  &  50.21  &  53.85 \\
\colrule
$K = \frac{\delta E/E}{\delta m_q/m_q}$
                & $-1.24$ & $-1.29$ & $-1.93$ & $-2.13$
                & $-1.50$ & $-1.49$ & $-1.59$ & $-1.70$
\end{tabular}
\end{ruledtabular}
\label{tab:av18uix}
\end{table*}

It is convenient to present the result for the variation
of the spin-orbit splitting in the following form:
\begin{equation}\label{deltaS}
\delta E_{so}=\delta E_{1/2}-\delta E_{3/2}=
E_{1/2}\frac{\delta E_{1/2}}{ E_{1/2}}-E_{3/2}\frac{\delta E_{3/2}}
{ E_{3/2}} \ .
\end{equation}
Accidentally, the calculated spin-orbit constant in  $^5$He is the same as
in $^{229}$Th, $V_{ls}=-0.83$ MeV (the $p_{1/2}$ - $p_{3/2}$ splitting
in $^5$He is $1.5 V_{ls}$). The spin-orbit constant in $^9$Be is larger
than in $^{229}$Th, in accord with the expected dependence $A^{-1/3}$
(see e.g. Ref.~\cite{BM}). The spin-orbit interval sensitivity
coefficients $K_{so}$ defined from 
\begin{equation}
  \frac{\delta E_{so}} {E_{so}}
   = K_{so} \frac{\delta m_q}{m_q}
\label{eq:so}
\end{equation}
for the quark mass variation in $^5$He, $^7$He, $^7$Li, and $^9$Be are 
$-0.27$, $-0.16$, $-2.58$, and $-0.22$, respectively.
The $^5$He, $^7$He, and $^9$Be values are all very similar, as all these
nuclei are essentially one nucleon outside a $0^+$ core.
The $^7$Li value is anomalously large because its ground and first
excited states are primarily a triton outside an alpha core, so
although $\delta E_{so}$ is comparable to $^9$Be, $E_{so}$ is very 
small and not typical of the single-particle spin-orbit interaction we seek.
Excluding the $^7$Li result gives us an average value of
$K_{so}=-0.22$ to use in $^{229}$Th.
Note that the estimate based on the Walecka model, outlined in Sec.~V
below, gives a very similar value $K_{so}=-0.2$.

\subsection{Frequency shift}

Substituting
  $\delta E_{so}/E_{so}=-0.22~\delta X_q/X_q$
  and  $\delta E_b/E_b = -1.45~\delta X_q/X_q$
into Eq.~(\ref{delta1}) we obtain the following energy shift
for the 7.6 eV transition in $^{229}$Th:
\begin{equation}\label{delta3}
\delta \omega=1.2~\frac{\delta X_q}{X_q}~{\rm MeV} \ .
\end{equation}
This corresponds to the frequency shift
$\delta \nu =3\cdot 10^{20}~\delta X_q/X_q$ Hz.
The width of this transition is $10^{-4}$ Hz so one may hope 
to get the sensitivity to the variation of $X_q$ about $10^{-25}$
per year. This is  $10^{11}$ times better than the current atomic clock
limit on the variation of $X_q$,  $\sim 10^{-14}$ per year
(see e.g. Ref.~\cite{F2007}).

The corresponding relative energy shift is
\begin{equation}\label{delta2}
  \frac{ \delta \omega}{\omega}=1.5 \cdot 10^5~\frac{\delta X_q}{X_q} \ .
\end{equation}
This enhancement coefficient may be compared with the
  coefficient $0.4 \cdot 10^5$ from
Ref.~\cite{th1} and $0.7 \cdot 10^5$ from
Ref.~\cite{He}.  The calculations in Ref.~\cite{He} have been done
using the relativistic mean field theory (extended Walecka model)
  and some basic ideas from Ref.~\cite{th1}. Thus, in this work
  we obtain an even larger enhancement! Here we present
the relative variations from Refs.~\cite{th1,He} for the new
  measured value 7.6 eV of
the frequency $\omega$ (the old value was 3.5 eV, and we 
multiplied  the numbers from \cite{th1,He} by  (3.5 eV)/$\omega$).
The difference between the results of different approaches
looks pretty large. However, this
is only a reflection of the current accuracy of all three calculations.
The present aim is to show that the enhancement does exist.

   Note that because of the huge enhancement it does not matter what
units one will use to measure the frequency $\omega$.
In the calculations above we assumed that $\omega$ is measured
in units of $\Lambda_{QCD}$. However, the variation
  of the ratio of any popular frequency standard to $\Lambda_{QCD}$
does not have such enhancement and  may be neglected.

\subsection{Coulomb energy and effect of  $\alpha$ variation}

We also would like to comment about the possible enhancement
of $\alpha$ variation. Ref.~\cite{Hayes} claims that
this enhancement is impossible since the ground and excited
states differ in the neutron state only and the neutron is neutral. 
Therefore, the ground and excited states have the same Coulomb energy 
and the interval does not change when $\alpha$ varies. 
We do not agree with this conclusion.
Indeed, the total Coulomb energy of the $^{229}$Th nucleus is 
$900$ MeV (see, e.g., \cite{BM}) which is
$10^8$ time larger than the energy difference $\omega$=7.6 eV.
Therefore, to have an enhancement it is enough to change 
the proton density distribution (deformation parameter)
by more than $10^{-8}$. Any change in the neutron state 
influences the nuclear mean field and proton distribution. 
For example, neutron removal changes the Coulomb energy of $^{229}$Th
by 1.3 MeV \cite{BM}.
This gives us an upper estimate (and a natural scale) for the
change of the Coulomb energy in the 7.6 eV $^{229}$Th transition.
One should expect a fraction of MeV change in any neutron
transition in heavy nuclei.  According to \cite{Gulda} the weight of
  admixed octupole  vibrations to the 7.6 eV state exceeds 20\%.
Octupole vibrations involve both protons and neutrons. Therefore, 
the proton density distribution  in the excited state is different
from the ground state and this difference is only an order of magnitude
smaller than the difference in neutron distribution.

The existence of the enhancement was confirmed by the direct
calculation  in Ref.~\cite{He}. The authors performed
the calculation of  the change of the
nuclear mean field acting on neutrons induced by the change
of the proton density due to the variation of $\alpha$. They obtained 
the enhancement coefficient $4 \cdot 10^3$. This corresponds to the
  Coulomb energy difference 0.03 MeV. As it was mentioned above,
an additional enhancement may come from the change of the nuclear
deformation. Anyway, there is no doubt that the enhancement 
of the sensitivity to $\alpha$ variation in $^{229}$Th
does exist. 

\section{Shift of the resonance in Samarium and  limits from the Oklo
natural nuclear reactor data}

In Refs.~\cite{Shuryak2002,Dmitriev} we derived a simple formula
to estimate the shift of the resonance or weakly bound energy level
  due to the variation  of the fundamental constants. Let us assume a
Fermi gas model in a  square well nuclear potential of the radius
$R$ and depth $V_0$. The energy of a single-particle energy level
or  resonance is determined as
\begin{equation}\label{res}
E_r \approx \langle \frac{p^2}{2m} \rangle -V_0 \ .
\end{equation}
The momentum $p$ in the square well is quantized, $p\approx {\rm constant}/R$.
Therefore, 
\begin{equation}\label{res1}
E_r=\frac{K}{2mR^2} -V_0 \ .
\end{equation} 
For a resonance or a weakly bound level $E_r\approx 0$
  ( $E_r \ll  V_0$) and
the constant $K \approx 2mR^2V_0$. Then we have 
\begin{equation}\label{res2}
\delta E_r=- \frac{K}{2mR^2}
  (\frac{ \delta m}{m}+ \frac{2 \delta R }{R}) - \delta V_0
  \approx -V_0 (\frac{ \delta m}{m}+ \frac{2 \delta R }{R}
  +\frac{ \delta V_0}{V_0} ) \ .
\end{equation} 
This equation is also valid for a compound state with several
excited particles. Indeed, the position of the compound state
or resonance relative to the bottom of the potential well
is determined mainly by the kinetic energy which scales
as $1/R^2$ (both the Fermi energy and sum of the single-particle
excitation energies scale this way). The shift of the resonance
due to the residual interaction between excited particles ($\sim 0.1$ MeV)
   is small in comparison with the depth of the potential well
($V_0 \approx 50$ MeV) and may be neglected.
  Note that the depth of the potential
$V_0$ is approximately the same  in light and heavy nuclei.
The radius of the well $R\approx 1.2 A^{1/3} r_0$, therefore,
the relative variation 
$\delta R/R = \delta r_0 /r_0$ is the same too. 
Thus, the resulting shift of the resonance both in light
  and heavy nuclei is given by Eq.~(\ref{res2}) and we may
  extrapolate the accurate result for light nuclei to
the resonance in $^{150}$Sm.

\begin{table}[ht!]
\caption{
}
\begin{ruledtabular}
\begin{tabular}{lrrrrrrrr}
& \multicolumn{1} {c}{$^5$He} & \multicolumn{1} {c}{$^6$He}
& \multicolumn{1} {c}{$^6$Li} & \multicolumn{1} {c}{$^7$He}
& \multicolumn{1} {c}{$^7$Li} & \multicolumn{1} {c}{$^7$Be}
& \multicolumn{1} {c}{$^8$Be} & \multicolumn{1} {c}{$^9$Be} \\
\colrule
$S_{\rm expt}$  & $-0.89$ &   1.86  &   4.59  & $-0.43$
                &   7.25  &   5.61  &  18.90  &   1.67 \\
$S_{\rm calc}$  & $-2.24$ & $-0.30$ &   2.96  & $-3.19$
                &   5.11  &   3.52  &  16.89  & $-3.24$\\
\colrule
$-\frac{\delta S_{\rm expt}}{\delta m_q/m_q}$
                &   3.42  &   9.92  &   9.52  &  11.73
                &  15.35  &  15.53  &  17.24  &  16.19 \\
$-\frac{\delta S_{\rm calc}}{\delta m_q/m_q}$
                &   1.62  &   6.12  &   7.06  &   4.58
                &  11.62  &  11.45  &  15.82  &   6.52 \\
\end{tabular}
\end{ruledtabular}
\label{tab:sep}
\end{table}

In Table~\ref{tab:sep} we present binding energies of the valence nucleon,
$S=-E$ (in MeV), and shift of the energy level (resonance),
$\frac{\delta E}{\delta m_q/m_q}=-\frac{\delta S}{\delta m_q/m_q}$, due to the
variation of the quark mass (in units MeV $\delta X_q/X_q$) in light
nuclei with $A=5,6,7,8,9$. In the derivation of Eq.~(\ref{res2}) it 
was assumed that the valence nucleon is localized inside the potential well.
This is not the case for $^5$He where the valence nucleon is localized
mainly outside the narrow potential well produced by the $^4$He core.
As a result the potential  $<V>$ averaged over the valence neutron
wave function $1p_{3/2}$ is significantly smaller than the depth
of the potential $V_0$. This explains why the shift in $^5$He (proportional
to $<V>$ - see Eq.~(\ref{res2}) and Ref.~\cite{Dmitriev}) is much smaller
than the shift in other nuclei. Another extreme case is $^8$Be where
$|E_r|$  is too large and the condition $E_r \ll V$ is not
fulfilled. The results for other nuclei are reasonably close to the
average value
\begin{equation}\label{shift}
\delta E_r \approx 10~\frac{\delta X_q}{X_q}~{\rm MeV} \ .
\end{equation}
We assume this shift for the 0.1 eV resonance in $^{150}$Sm.
This value does not contradict the order-of-magnitude estimates
in Refs.~\cite{Shuryak2002,Dmitriev,F2007}. Finally, we can add to this
shift the contribution of $\alpha$ variation  from Refs.~\cite{Dyson}
($\delta E_r=-1.1 \pm 0.1$ MeV $\delta \alpha /\alpha$).
The total shift of the resonance in $^{150}$Sm is
\begin{equation}\label{shift1}
\delta E_r=10~(\frac{\delta X_q}{X_q}-0.1 \frac{\delta \alpha}{\alpha})~{\rm MeV} \ .
\end{equation}

   Now we can can extract limits on the variation of $X_q$ from the
measurements of $\delta E_r$. Pioneering work in this area was
  done in Ref.~\cite{Dyson}.
We will use recent measurements \cite{Gould,Petrov,Fujii}
  where the accuracy is higher. Ref.~\cite{Fujii}
  has given $|\delta E_r|< 20$ meV. Then Eq.~(\ref{shift1}) gives 
\begin{equation}\label{limit1}
|\frac{\delta X_q}{X_q}-0.1 \frac{\delta\alpha}{\alpha}| < 2 \cdot 10^{-9} \ .
\end{equation}
Ref.~\cite{Petrov}  has given -73 $<\delta E_r< $ 62 meV. This gives
\begin{equation}\label{limit2}
|\frac{\delta X_q}{X_q}-0.1 \frac{\delta\alpha}{\alpha}| < 7 \cdot 10^{-9} \ .
\end{equation}
  Ref.~\cite{Gould}  has given -11.6 $<\delta E_r< $ 26.0 meV. This gives
\begin{equation}\label{limit3}
|\frac{\delta X_q}{X_q}-0.1 \frac{\delta\alpha}{\alpha}| < 2.6 \cdot 10^{-9} \ .
\end{equation}
The limits on  $\delta E_r$ have been been presented with
  $2\sigma$ range. Note that Ref.~\cite{Gould} has presented
also the second, non-zero solution
  (it exists since the resonance has two tails):
-101.9 $<\delta E_r< $ -79.6 meV.  However, Ref.~\cite{Fujii} 
tentatively ruled out this solution based on the data
for the shift of a similar resonance in the Gd nucleus.

  Based on the results above we conclude  that
  $|\frac{\delta X_q}{X_q}|< 4 \cdot 10^{-9}$ (for simplicity, we omit the
small contribution of $\alpha$ variation here). Assuming linear time
dependence during the last 1.8 billion years we obtain the best terrestrial
  limit on the variation of the fundamental constants
\begin{equation}\label{limit4}
|\frac{{\dot X}_q}{X_q}| < 2.2 \cdot 10^{-18} y^{-1} \ .
\end{equation}

\section{Variation of nuclear radius}
Variation of the nuclear radius is needed to calculate effects of the 
fundamental constant variation in microwave atomic clocks
where the transition frequency depends on a probability
of the electron to be inside the nucleus. 
Indeed, the hyperfine interaction constant in heavy atoms has some
 sensitivity to  the nuclear radius (including the Cs hyperfine transition
 which defines the unit of time, the second, and is used as a reference
in numerous atomic and molecular clock experiments). This dependence
was also requested by S. Schiller who proposed new experiments
with hydrogen-like ions  to search for the variation of the fundamental
 constants \cite{Schiller}.

In  Table \ref{tab:radii} we present a comparison of calculated
and measured charge nuclear radii for the stable $A=2,7$ nuclei.
Determination of the sensitivity of the nuclear radius to quark mass
variation is a more involved calculation than for the energy.
While the deuteron can be solved exactly, the VMC calculations for
$A \ge 3$ nuclei of Ref.~\cite{FW07} have to be modified.
This is because the variational bound for the energy is a quadratic 
function near its minimum in the space of variational parameters, but the 
radius is a linear function.
In the previous VMC calculations, the variational parameters were fixed
at the energy minimum for the nominal hadron masses corresponding to 
$\delta m_q = 0$, and then not allowed to vary as the energy was evaluated
for different $\delta m_H$.
Consequently the ``size" of the trial wave function was essentially unchanged.
For the radius determination, we must allow this size to vary.
We do this by multiplying a set of variational parameters (those to 
which the radius is most sensitive) by a scale factor, and then carefully 
reminimize this scale factor for each $\delta m_H$.
This allows us to determine 
$\Delta r(m_H) = \frac{\delta r/r}{\delta m_H/m_H}$.
The $\Delta{\cal E}(m_H) = \frac{\delta E/E}{\delta m_H/m_H}$ 
reported in Ref.~\cite{FW07} are unchanged in this new minimization.

\begin{table}[ht!]
\caption{Experimental and calculated point proton rms radii for stable
$A=2-9$ nuclei.}
\begin{ruledtabular}
\begin{tabular}{lrrrrrrrrr}
& \multicolumn{1} {c}{$^2$H}  & \multicolumn{1} {c}{$^3$H}
& \multicolumn{1} {c}{$^3$He} & \multicolumn{1} {c}{$^4$He}
& \multicolumn{1} {c}{$^6$He} & \multicolumn{1} {c}{$^6$Li}
& \multicolumn{1} {c}{$^7$Li} & \multicolumn{1} {c}{$^7$Be} 
& \multicolumn{1} {c}{$^9$Be} \\
\colrule
 AV18+UIX  & 1.967 & 1.58 & 1.77 & 1.45 & 1.92 & 2.46 & 2.34 & 2.45 & 2.40 \\
 Expt.     & 1.953 & 1.59 & 1.75 & 1.45 & 1.93 & 2.39 & 2.25 &      & 2.38 \\
\end{tabular}
\end{ruledtabular}
\label{tab:radii}
\end{table}

This procedure works well for the $A=3,4$ nuclei, and the $\Delta r(m_H)$
is presented in Table~\ref{tab:deltar}, along with the total sensitivity
to the quark mass, $K_r \equiv \frac{\delta r/r}{\delta m_q/m_q}$,
obtained by folding in the DSE values for $\delta m_H/m_H$.
However, because our trial functions for $A=6-9$ nuclei are not
inherently stable against breakup into subclusters, we need to make 
an additional constraint when calculating their sensitivity.
For some of these nuclei, we have trial functions that asymptotically
look like the appropriate subclusters bound in a Coulomb well with
the experimental separation energy: $^6$Li is asymptotically an alpha
and a deuteron bound by 1.47 MeV, $^7$Li is asymptotically an alpha
and a triton bound by 2.47 MeV, and $^7$Be is asymptotically an alpha
and a $^3$He bound by 1.59 MeV.
($^6$He is asymptotically a three-body $\alpha+n+n$ cluster and $^9$Be
is an $\alpha+\alpha+n$ cluster, so they cannot be treated this way.)
For a quark mass shift $\delta m_q/m_q=\pm 0.01$, we know the total
energy shift expected from our previous calculations.
We subtract that portion attributable to the alpha and deuteron or
trinucleon subclusters, and use the remaining energy shift to adjust
the asymptotic separation energy of our trial function.
This allows the size of both the subclusters and the well binding
them to vary.
For $A=6,7$, we have carried out this calculation for the total
sensitivity $K_r$ only, and not for the individual $\Delta r(m_H)$; 
these results are also given in Table~\ref{tab:deltar}.

\begin{table}[ht!]
\caption{Dimensionless derivatives of point proton rms radii
$\Delta{r}(m_H) = \frac{\delta r/r}{\delta m_H/m_H}$ and the
sensitivity with respect to $m_q$ after folding in the DSE values of
$\delta m_H/m_H$.}
\begin{ruledtabular}
\begin{tabular}{lrrrrrrr}
& \multicolumn{1} {c}{$^2$H}  & \multicolumn{1} {c}{$^3$H}
& \multicolumn{1} {c}{$^3$He} & \multicolumn{1} {c}{$^4$He}
& \multicolumn{1} {c}{$^6$Li} & \multicolumn{1} {c}{$^7$Li}
& \multicolumn{1} {c}{$^7$Be} \\
\colrule
$m_N+\delta_N$  & $-7.32$ & $-4.81$ & $-4.73$ & $-3.04$ &       &      & \\
$\delta_\Delta$ &   4.07  &   3.32  &   3.28  &   2.18  &       &      & \\
$m_\pi$ (+TNI)  &   2.57  &   1.80  &   1.77  &   1.11  &       &      & \\
$m_V$           &$-16.39$ &$-12.97$ &$-12.79$ & $-8.50$ &       &      & \\
\colrule
$\frac{\delta r/r}{\delta m_q/m_q}$
                &   0.48  &   0.34  &   0.33  &   0.20  & 0.35  & 0.27 & 0.22 \\
\end{tabular}
\end{ruledtabular}
\label{tab:deltar}
\end{table}

The average value of $K_r$ is about 0.3, which may serve as an estimate
of the sensitivity for all nuclei.
There are significant deviations from this value for the very weakly bound
deuteron $^2$H and very strongly bound $^4$He; the latter is probably a
solid lower bound.

    The dependence of the nuclear radius on fundamental constants manifests
itself in microwave transitions in atomic clocks which are used to search
for the variation of the fundamental constants 
(see e.g. Refs.~\cite{F2007,Schiller}).
The dependence of the hyperfine transition frequency $\omega_h$ on nuclear
radius $r$ in atoms with an external $s$-wave electron is approximately given 
by the following expressions (in units of $\Lambda_{QCD}$):
\begin{equation}\label{hyperfine1}
\frac{\delta \omega_h}{\omega_h}= K_{hr} \frac{\delta r}{r}
=K_{hr} K_r \frac{\delta m_q}{m_q}
 \approx 0.3~K_{hr} \frac{\delta m_q}{m_q} \ ,
\end{equation}      
\begin{equation}\label{hyperfine2}
 K_{hr} \approx - 
\frac{(2 \gamma-1) \delta_h}{1-\delta_h} \ ,
\end{equation}  
\begin{equation}\label{hyperfine3}
 \delta_h \approx 2~(3 \cdot 10^{-5} Z^{4/3})^{2 \gamma-1} \ ,
\end{equation}  
where $\gamma=(1-Z^2\alpha^2)^{1/2}$. For the Cs atom microwave standard
the nuclear charge $Z=55$ and  $K_{hr}=-0.03$; for the Hg$^+$ microwave clock
 $Z=80$ and  $K_{hr}=-0.09$.

  We also calculated the dependence of the $^4$He radius on $\alpha$:
$\frac{\delta r/r}{\delta \alpha/\alpha}=0.0034$. For heavy nuclei
the relative role of the Coulomb repulsion increases and the sensitivity
to the $\alpha$ variation should be larger.

\section{Estimates in Walecka model}

It is instructive to compare the results obtained by the extrapolation
from light nuclei with some ``direct'' calculations.
In this section we estimate the variations of the resonance
positions and spin-orbit splittings in heavy nuclei using
the Walecka model \cite{SW} where the strong nuclear potential
is produced by scalar and vector meson exchanges:
\begin{equation}\label{walecka}
V= -{g_s^2 \over 4\pi} {e^{-r m_S} \over r}+{g_v^2\over 4\pi}
  {e^{-r m_V} \over r} \ .
\end{equation} 
Averaging Eq.~(\ref{walecka}) over the nuclear volume
  we can find the depth of the potential well
  \cite{Shuryak2002}
\begin{equation}\label{depth}
  V_0= \frac{3}{4 \pi r_0^3} \left(\frac{g_s^2}{m_S ^2} -
\frac{g_v^2}{m_V ^2} \right) \ .
\end{equation}
  Here $2r_0 = 2.4$ fm is an internucleon distance.
The result for the variation of the potential is 
\begin{equation}{\delta V_0 \over V_0} \approx - 7.5 \frac{\delta m_S}{m_S}+
  5.5 \frac{\delta m_V}{m_V}-3 \frac{\delta r_0}{r_0} \ .
\end{equation}
Here we have used 
$\frac{g_s^2}{m_S ^2}/\frac{g_v^2}{m_V ^2} = 266.9/195.7 = 1.364$ from
 Ref.~\cite{C}. There is an order of magnitude enhancement of the meson mass
 variation contributions due to the
 cancellation of the vector and scalar contributions in the denominator $V_0$.
 Eq.~(\ref{res2}) for the variation of the resonance position becomes
\begin{equation}\label{res3}
\delta E_r \approx V_0 ( 7.5 \frac{\delta m_S}{m_S}- 5.5 \frac{\delta m_V}{m_V}
-\frac{ \delta m_N}{m_N}+ \frac{\delta r_0 }{r_0} ) \ .
\end{equation} 
We do not know the variation of $r_0$ in the Walecka model, therefore, to make
a rough numerical estimate we neglect this term.
As above we take  dependence
of the nucleon and  meson masses on the  current light quark
   mass $m_q=(m_u+m_d)/2$ from  Refs \cite{FHJRW06,HMRW}:
$\frac{\delta m_{\omega}}{m_{\omega}} = 0.034 \frac{\delta m_q}{m_q }$,
$\frac{\delta m_N}{m_N} = 0.064 \frac{\delta m_q}{m_q }$,
$\frac{\delta m_{\sigma}}{m_{\sigma}} = 0.013 \frac{\delta m_q}{m_q }$,
$\frac{\delta m_{\pi}}{m_{\pi}} = 0.498 \frac{\delta m_q}{m_q }$.
The vector meson in the Walecka model is usually identified
with the $\omega$-meson so
 $\frac{\delta m_V}{m_V} = 0.034 \frac{\delta m_q}{m_q }$. 
The scalar meson exchange, in fact, imitates both the
 $\sigma$ meson exchange and two-pion exchange.
Even if we neglect the two-pion exchange in zero approximation,
there is virtual $\sigma$ decay to two $\pi$. These virtual decays
(loops on $\sigma$ line in the NN-interaction diagrams with intermediate
$\sigma$) very strongly modify the $\sigma$ propagator and change its
 large distance asymptotics from $e^{-m_{\sigma}r}$ to  $e^{-2 m_{\pi}r}$
\cite{Shuryak2007}. The  mixing between $\sigma$ and 
two $\pi$ in $m_S$ should increase the sensitivity coefficient for
the variation of $m_S$. For an estimate we take an intermediate
value between the neutron and vector meson mass sensitivity,
 $\frac{\delta m_S}{m_S} \sim 0.05 \frac{\delta m_q}{m_q }$.
(Note that the  positive contribution of $\frac{\delta r_0 }{r_0}$
in Eq.~(\ref{res3}) produces an effect similar to that of an increase
of $\frac{\delta m_S}{m_S}$.)
Then Eq.~(\ref{res3}) gives
 \begin{equation}\label{shiftW}
\delta E_r \sim 10~\frac{\delta X_q}{X_q}~{\rm MeV} \ .
\end{equation}
This rough estimate agrees with the result extrapolated from light nuclei.
Note, however, that the accuracy of this estimate is very low due to
the cancellations of different terms.

The scalar and vector mesons contribute with
  equal sign to the spin-orbit interaction constant $V_{ls}$ \cite{C}.
Also,  the spin-orbit interaction
is inversely proportional to the nucleon mass $m_N$  squared.
  Thus, we have
\begin{equation}
  V_{ls} \propto \frac{1}{m_N^2}
\left(\frac{g_s^2}{m_S ^2} + \frac{g_v^2}{m_V ^2} \right) \ ,
\end{equation}
\begin{equation}
\frac{\delta E_{so}}{E_{so}}=-2~(\frac{\delta m_N}{m_N}+
\frac{0.58\delta m_S}{m_S}+
\frac{0.42\delta m_V}{m_V})\approx -0.2~\frac{\delta m_q}{m_q } \ .
\end{equation}
This estimate is close to the result ($-0.22~\frac{\delta m_q}{m_q }$)
 obtained by the extrapolation from light nuclei. Note, however, that here
we neglected the effect of variation of $r_0$ which probably should
increase the absolute value of the sensitivity coefficient.

\section{Conclusion}

At the moment one can hardly calculate the sensitivity coefficient for the 
dependence of the strong interaction on the quark mass $m_q$ with an accuracy 
better than a factor of 2.  Moreover, it is hard to identify this dependence
in phenomenological interactions which are used for the calculations
in heavy nuclei. For example, it is not obvious that the  scalar and  vector
mesons in the Walecka model are actually equivalent to free $\sigma$
and  $\omega$ mesons in particle physics. Therefore, to test conclusions
obtained using the Walecka model, we explored a complementary approach.
We performed the calculations in light nuclei where the interactions
are well-known and the accuracy of the calculations is high.
  The binding energy per nucleon $E_b$, the spin-orbit interaction constant
$V_{ls}$ and the nuclear radius $r$ have a slow dependence  as a function of
 the nucleon number $A$. Moreover,
the common factors (like $A^{-1/3}$ in the spin-orbit constant $V_{ls}$
and  $A^{1/3}$ in the nuclear radius)
  cancel out in the relative variations $\delta r/r$,
 $\delta V_{ls}/ V_{ls}$
and $\delta E_b/ E_b$. Therefore, we can extract these relative
variations from the calculations in light nuclei and use them in heavy nuclei.
The errors produced by such extrapolation may be
smaller than the errors of direct calculations in heavy nuclei.
So far, this extrapolation
and direct calculations using Walecka model
  give comparable values of the enhancement factors in $^{229}$Th and
  $^{150}$Sm.

\acknowledgments

VVF is grateful to H. Feldmaier for useful discussions.
This work is supported by the U.S.\ Department of Energy, Office of
Nuclear Physics, under contract DE-AC02-06CH11357, and by
the Australian Research Council.
Calculations were made at Argonne's Laboratory Computing Resource Center.


\begin{thebibliography}{99}

\bibitem{Uzan}
J-P. Uzan, Rev. Mod. Phys. {\bf 75}, 403 (2003).

\bibitem{F2007} 
V. V. Flambaum, Int. J. Mod. Phys. A {\bf 22}, 4937 (2007).

\bibitem{th1}
V. V. Flambaum,  Phys. Rev. Lett. {\bf 97}, 092502 (2006).

\bibitem{th4}
E. Peik and Chr. Tamm, Europhys. Lett. {\bf 61}, 181 (2003).

\bibitem{th2}
E. V. Tkalya, A. N. Zherikhin, and V. I. Zhudov, 
Phys. Rev. C {\bf 61}, 064308 (2000);
A. M. Dykhne, E. V. Tkalya, 
Pis'ma Zh. Eks. Teor. Fiz. {\bf 67}, 233 (1998)
[JETP Lett. {\bf 67}, 251 (1998)].

\bibitem{Beck} 
B. R. Beck, J. A. Becker, P. Beiersdorfer, G. V. Brown, K. J. Moody, 
J. B. Wilhelmy, F. S. Porter, C. A. Kilbourne, and R. L. Kelley,
Phys. Rev. Lett. {\bf 98}, 142501 (2007).

\bibitem{th6} 
Z. O. Guimar\~{a}es-Filho and O. Helene, Phys. Rev. C {\bf 71}, 044303 (2005).

\bibitem{th} 
R. G. Helmer and C. W. Reich, Phys. Rev. C {\bf 49}, 1845 (1994).

\bibitem{exp} E. Peik. Talk at  workshop ``In search for variation of
 fundamental constants and mass scales'',Perimeter Institute, July14-18, 2008.
 E. Hudson, Talk at  workshop ``In search for variation of fundamental
 constants and mass scales'', Perimeter Institute, July14-18, 2008.
D. Habbs, privite communication. D.DeMille, privite communication.
 J. Torgerson, privite communication.

\bibitem{Hayes} 
A. C. Hayes and J. L. Friar, Phys. Lett. B {\bf 650}, 229 (2007).

\bibitem{Gould}
C. R. Gould, E. I. Sharapov, and S. K. Lamoreaux,
Phys. Rev. C {\bf 74}, 024607 (2006).

\bibitem{Petrov}
Yu. V. Petrov, A. I. Nazarov, M. S. Onegin, V. Yu. Petrov, and E. G. Sakhnovsky,
Phys. Rev. C {\bf 74}, 064610 (2006).

\bibitem{Fujii}
Y. Fujii, A. Iwamoto, T. Fukahori, T. Ohnuki, M. Nakagawa, H. Hidaka, Y. Oura,
and P. M\"{o}ller, Nucl. Phys. {\bf B573}, 377 (2000).

\bibitem{Marciano}
W. J. Marciano, Phys. Rev. Lett. {\bf 52}, 489 (1984);
X. Calmet and H. Fritzsch, Eur. Phys. J. {\bf C24}, 639 (2002);
P. Langacker, G. Segr\'{e}, and M. J. Strassler, Phys. Lett. {\bf B528}, 121
(2002);
T. Dent and M. Fairbairn. Nucl. Phys. {\bf B653}, 256 (2003);
C. Wetterich, JCAP {\bf 10}, 002 (2003);  Phys. Lett. {\bf B561}, 10 (2003).

\bibitem{Shuryak2003} 
V. V. Flambaum and E. V. Shuryak, Phys. Rev. D {\bf 67}, 083507 (2003).

\bibitem{BM} 
A. Bohr and B. R. Mottelson, {\it Nuclear Structure Volume II},
(W. A. Benjamin, New York, 1974).

\bibitem{Gulda} 
K. Gulda {\it et al.}, Nucl. Phys. {\bf A703}, 45 (2002). 

\bibitem{C} 
R. Brockmann and W. Weise, Phys. Rev. C {\bf 16}, 1282 (1977).

\bibitem{FW07}
V. V. Flambaum and R. B. Wiringa, Phys. Rev. C {\bf 76}, 054002 (2007).

\bibitem{PP93}
S. C. Pieper and V. R. Pandharipande, Phys. Rev. Lett. {\bf 70}, 2541 (1993).

\bibitem{FHJRW06}
V. V. Flambaum, A. H\"{o}ll, P. Jaikumar, C. D. Roberts, and S. V. Wright,
Few-Body Syst. {\bf 38}, 31 (2006).

\bibitem{HMRW} 
A. H\"{o}ll, P. Maris, C. D. Roberts, and S. V. Wright,
arXiv:nucl-th/0512048v1. 

\bibitem{He} Xiao-tao He and Zhong-zhou Ren, J. Phys. G: Nucl. Part. Phys. {\bf 34}, 1611 (2007).

\bibitem{Shuryak2002}
V. V. Flambaum and E. V. Shuryak, Phys. Rev. D {\bf 65}, 103503 (2002).

\bibitem{Dmitriev}
V. F. Dmitriev and V. V. Flambaum, Phys. Rev. D {\bf 67}, 063513 (2003).

\bibitem{Dyson} 
A. I. Shlyakhter, Nature (London), {\bf 264}, 340 (1976); 
Yu. V. Petrov, Sov. Phys. Usp.  {\bf 20}, 937 (1977); 
T. Damour and F. Dyson, Nucl. Phys. {\bf B480}, 37 (1996).

\bibitem{Schiller} S. Schiller,  Phys. Rev. Lett. {\bf 98}, 180801 (2007).

\bibitem{SW} B. D. Serot and J. D. Walecka,
Adv. Nucl. Phys. {\bf 16}, 1 (1986).

\bibitem{Shuryak2007} 
V. V. Flambaum and E. V. Shuryak, Phys. Rev. C  {\bf 76}, 065206 (2007).

\end{thebibliography}
\end{document}